\documentclass[12pt]{iopart}

\usepackage{iopams}
\usepackage{graphicx}

\begin{document}

\title[Generalized Langevin Equation Formulation for Anomalous Polymer
  Dynamics]{Generalized Langevin Equation Formulation for Anomalous Polymer
  Dynamics}

\author{Debabrata Panja}
\address{Institute for Theoretical Physics, Universiteit van
Amsterdam, Valckenierstraat 65, 1018 XE Amsterdam, The Netherlands}
\eads{\mailto{D.Panja@uva.nl}}

\begin{abstract} 
For reproducing the anomalous --- i.e., sub- or super-diffusive ---
behavior in some stochastic dynamical systems, the Generalized
Langevin Equation (GLE) has gained considerable popularity in recent
years. Motivated by the question whether or not a system with
anomalous dynamics can have the GLE formulation, here I consider
polymer physics, where sub-diffusive behavior is commonplace. I
provide an exact derivation of the GLE for phantom Rouse polymers,
and by identifying polymeric response to local strains, I argue the
case for the GLE formulation for self-avoiding polymers and polymer
translocation through a narrow pore in a membrane. The number of
instances in polymer physics, where the anomalous dynamics
corresponds to the GLE, thus seems to be fairly common.
\end{abstract}


\maketitle

Following the work of Einstein and Smoluchowski, a century ago
Langevin proposed a stochastic dynamical description of Brownian
motion \cite{langevin}. The corresponding Langevin equation (LE) for a
particle of mass $m$ and velocity $v(t)$ in a fluid of damping
coefficient $\gamma$ (meaning that the friction coefficient is
$m\gamma$) is given by
\begin{eqnarray}
m\dot v(t) = -m\gamma v(t) + f(t),
\label{e1}
\end{eqnarray} 
where the stochastic force $f(t)$ satisfies $\langle f(t)\rangle=0$
and the fluctuation-dissipation theorem (FDT) $\langle
f(t)f(t')\rangle=2m\gamma k_BT\delta(t-t')$. Here $k_B$ is the
Boltzmann constant, $T$ is the absolute temperature of the fluid and
the angular brackets denote equilibrium ensemble averaging. Having
found wide-ranging applications \cite{lang} the LE (\ref{e1}) ---
describing the Brownian motion in the so-called Rayleigh formulation
--- has long since been incorporated into the fundamentals of
stochastic processes \cite{vankampen,kubo}. An exact result of the LE
is that the dynamics of the Brownian particle is diffusive at long
times, with the diffusion coefficient $D=k_BT/(m\gamma)$
\cite{einstein}. However, there is a diverse range of physical
processes, for which the diffusive behavior at long times is not the
norm; instead the mean-square-displacement (MSD) of the particle is
anomalously fast or slow, increasing in time as $t^\alpha$ for
$\alpha\neq1$. For describing the dynamics in some of these systems,
e.g., turbulent diffusion \cite{bouch}, disorder related excitations
\cite{moura}, ATP coupling to motor proteins \cite{bier}, dipolar
chains in a ferrofluid \cite{ferro1}, ferrofluid patterns in a
magnetic field \cite{ferro2}, and traffic flows \cite{traffic}, a
generalization of the LE in the Rayleigh formulation (\ref{e1}) ---
the so-called Generalized Langevin Equation (GLE) of the Mori-Lee form
\cite{morilee} --- given by
\begin{eqnarray} 
m\dot v(t)=-m\int_0^t dt'\, \Gamma(t-t')\, v(t')+g(t)
\label{e2}
\end{eqnarray}
has gained considerable popularity in recent years. Here, the
stochastic force $g(t)$ satisfies $\langle g(t)\rangle=0$ and the
corresponding  FDT $\langle g(t) g(t')\rangle=m
k_BT\,\Gamma(t-t')$. The GLE reduces to the LE when
$\Gamma(t-t')=2\gamma\delta(t-t')$. It is the non-instantaneous nature
of $\Gamma(t)$ in the GLE that leads to the anomalous dynamics: the
result that if $\Gamma(t)\sim t^{-\alpha}$ for some $\alpha$ at long
times, then the  particle's MSD $\sim t^\alpha$, has been derived not
so long ago \cite{oliprl}.

The characterization of the anomalous dynamics, given the
characteristics of $\Gamma(t)$ in the GLE, is a one-way street. From
this perspective, it  would also be worthwhile to know whether or not
a system with anomalous  dynamics can have the GLE formulation. While
a generic answer to this  question is not known to the best of my
knowledge, I note that the  complexity of the systems that exhibit
anomalous dynamics typically  presents a barrier for the
answer. Motivated by this question, I consider in this Letter polymer
physics, where anomalous dynamics is  commonplace. I derive the GLE
exactly for the motion of a tagged  monomer for a phantom Rouse
polymer. [The phantom Rouse equation  follows the so-called
Smoluchowski formulation \cite{de}, hence the  corresponding GLE is
not of the Mori-Lee form (\ref{e2}); see later].  Further, I identify
polymeric response to local strains, and argue the  existence of the
GLE for (single) self-avoiding polymers and polymer
translocation. Interestingly, despite the fact that any first course
on polymer physics teaches that the dynamics of a tagged monomer of a
polymer is anomalous until the terminal time $\tau$ (the relaxation
time of the polymer), I have not seen, in published literature, the
derivation of the GLE for polymer dynamics, and subsequently, the
follow-up to the anomalous dynamics. This Letter thus provides the
first indication that the number of instances in polymer physics,
where the anomalous dynamics corresponds to the GLE, may indeed be
fairly common. Whether or not this procedure for the GLE formulation
can be extended to other systems (that exhibit anomalous dynamics) is
also brought into consideration.

The correct single polymer dynamics in a fluid was first presented by
Zimm \cite{zimm}. I refer to such polymers, for which the monomers
interact with each other via hydrodynamic interactions, in this
Letter, as Zimm polymers. Few years earlier than Zimm, Rouse
\cite{rouse} put forward a model for single polymer dynamics that
neglect the hydrodynamic interactions between the monomers; although
incorrect, the corresponding (Rouse) polymer dynamics resides at the
heart of polymer physics \cite{de} --- and widely used till today ---
largely due to its simplicity.

{\it The GLE for phantom Rouse polymers.} Consider a phantom Rouse
polymer (of length $N$); i.e., a Rouse polymer that can intersect
itself. It is described by the Rouse equation; in continuum
representation it reads \cite{de}
\begin{eqnarray}  
\gamma\frac{\partial\vec r_n}{\partial t}=k\, \frac{\partial^2 \vec
r_n(t)}{\partial n^2}+\vec f_n(t),
\label{e3}
\end{eqnarray}   
where $\vec r_n(t)$ is the location of the $n$-th monomer at time $t$,
$\gamma$ is the damping coefficient of the surrounding fluid, and $k$
is the spring constant for the springs connecting the consecutive
monomers. The stochastic force $\vec f_m(t)$ satisfies the conditions
$\langle\vec f_n(t)\rangle=0$ and $\langle
f_{m\kappa}(t)f_{n\lambda}(t')\rangle=2\gamma
k_BT\delta(m-n)\delta_{\kappa\lambda}\delta(t-t')$, for
$\kappa,\lambda=(x,y,z)$. Equation (\ref{e3}) is supplemented by the
``open'' boundary conditions that the chain tension of the polymer at
the free ends must vanish; i.e., $(\partial\vec r_n/\partial
n)|_{n=0}=(\partial\vec r_n/\partial n)|_{n=N}=0$.

Since the Rouse equation is linear in $\vec r_n(t)$, it can be solved
to obtain all correlation functions using the mode expansion technique
\cite{de}. Two noteworthy results borne out of this exercise are: (a)
the terminal relaxation time $\tau=\gamma N^2/(\pi^2k)$, and (b) the
MSD of the middle monomer increases as $t^{1/2}$ until time $\tau$,
and only after that time the middle monomer performs diffusive motion,
with the diffusion coefficients scaling as $1/N$. It is this
sub-diffusive motion of the middle monomer that I obtain from the GLE,
which in turn I derive exactly from Eq. (\ref{e3}). More precisely, I
demonstrate that the chain tension $\vec\phi(t)$ that the middle
monomer experiences is obtained from the velocity of the middle
monomer $\vec v(t)$ via the GLE
\begin{eqnarray}  
\vec\phi(t)=-\int_0^t dt'\, \mu(t-t')\, \vec v(t')+\vec g(t),
\label{egleb}
\end{eqnarray}  
with $\vec g(t)\equiv\vec\phi(t)_{\vec v=0}$, so that $\langle\vec
g(t)\rangle=0$, and $\langle\vec g(t)\cdot\vec
g(t')\rangle=3\,k_BT\,\mu(t-t')=6k_BT\,\sqrt{\pi\gamma
k}(t-t')^{-1/2}\exp[-(t-t')/\tau]$ is the FDT. The factor $3$ in the
expression for $\langle\vec g(t)\cdot\vec g(t')\rangle$ stems from the
fact that I consider polymers in three dimensions. From Eq.
(\ref{egleb}) I further arrive at the GLE
\begin{eqnarray}  
\vec v(t)=-\int_0^t dt'\, a(t-t')\,\vec\phi(t')+\vec h(t),
\label{eglea}
\end{eqnarray}  
which encodes the anomalous dynamics of the middle monomer. Here
$\langle\vec h(t)\rangle=0$, $\langle\vec h(t)\cdot\vec
h(t')\rangle=3k_BT\,a(t-t')$ is the FDT, and $\tilde\mu(s)\tilde
a(s)=1$ in the Laplace space.

Two ingredients are necessary to derive Eq. (\ref{egleb}). The first
one of them is the dynamics for both halves of the polymer when the
middle monomer held fixed at, say, $\vec {\cal R}$. With $\vec
r\,'_n(t)=\vec r_n(t)-\vec{\cal R}$, I define
\begin{eqnarray}   
\vec Y^{(r)}_p(t)=\frac1N\int_{0}^{\frac N2}dn\,\sin\frac{\pi
(2p+1)n}N\, \vec r\,'_{n+N/2}(t),
\label{e6}
\end{eqnarray} 
\begin{eqnarray}  
\mbox{and}\,\,\vec Y^{(l)}_p(t)=-\frac1N\int_{-\frac N2}^0 dn\,
\sin\frac{\pi (2p+1)n}N\, \vec r\,'_{n+N/2}(t)
\label{e7}
\end{eqnarray} 
for $p=0,1,\ldots$, for the right and the left half, such that
\begin{eqnarray} 
\vec r\,'_{n+N/2}(t)=4\sum_p\vec
Y^{(r)}_p(t)\sin\frac{\pi(2p+1)n}N\,\Theta(n)\nonumber\\&&\hspace{-5.6cm}-4\sum_p\vec
Y^{(l)}_p(t)\sin\frac{\pi(2p+1)n}N\,\Theta(-n).
\label{e10}
\end{eqnarray}  
Equation (\ref{e10}) shows that the open boundary conditions are
satisfied, while the middle monomer remains fixed. The {\it
independent\/} evolution of each half satisfies
\begin{eqnarray} 
\gamma_p\frac{\partial\vec Y_p}{\partial t}=-k_p\,\vec Y_p(t)+\vec
f_p(t),
\label{e9}
\end{eqnarray} 
for $\vec Y_p=[\vec Y^{(l)}_p,\vec Y^{(r)}_p]$, $\gamma_p=2N\gamma$,
$k_p=2\pi^2k(2p+1)^2/N$, $\langle\vec f_p\rangle=0$ and $\langle
f_{p\kappa}(t)f_{q\lambda}(t')\rangle=\delta_{pq}\delta_{\kappa\lambda}\gamma_pk_BT\delta(t-t')$.

The second ingredient is the equilibrium averages of $Y_p$  and
$Y_p^2$. First, $\langle\vec Y^{(r)}_p(t)\rangle=\langle \vec
Y^{(l)}_p(t)\rangle\equiv0$ by isotropy. Secondly, by left-right
symmetry
$\langle[Y^{(r)}_p(t)]^2\rangle=\langle[Y^{(l)}_p(t)]^2\rangle$;  and
they are obtained from the LE (\ref{e9}) as
\begin{eqnarray}
\hspace{-2mm}\langle[Y^{(r)}_p(t)]^2\rangle=\langle[Y^{(l)}_p(t)]^2\rangle=
3Nk_BT/[4\pi^2k(2p+1)^2].
\label{e8}
\end{eqnarray}

With these ingredients, I now follow the dynamics of the middle
monomer. Consider the case where the middle monomer of a polymer,
stationary at $t=0^-$, moves by a distance $\vec{\delta r}_0$ at
$t=0$, corresponding to $\vec v(t)=\vec{\delta r}_0\delta(t)$. Then,
for $\vec Y_p=[\vec Y^{(l)}_p,\vec Y^{(r)}_p]$, I have
\begin{eqnarray} 
\vec Y_p(0^+)=\vec Y_p(0^-)-\vec{\delta r}_0/[\pi(2p+1)]
\label{e11}
\end{eqnarray} 
Until the time the middle monomer moves again, it feels
$k(\partial\vec r\,'_{n+N/2}/\partial n)_{n=0^+}$ and $-k(\partial\vec
r\,'_{n+N/2}/\partial n)_{n=0^-}$ forces from the right and the left
half of the polymer respectively. The total force $\vec\phi(t)$
experienced by the middle monomer is their sum, and from Eq.
(\ref{e10}) it is given by
\begin{eqnarray} 
\vec\phi(t)=4k\sum_p\frac{\pi(2p+1)}N\left[\vec Y^{(r)}_p(t)+\vec
Y^{(l)}_p(t)\right].
\label{e12}
\end{eqnarray} 
To obtain the time evolution of $\vec\phi(t)$ until the middle monomer
moves again, we return to Eq. (\ref{e9}) and write
\begin{eqnarray} 
\vec Y_p(t)\!=\!e^{-k_pt/\gamma_p}\vec
Y_p(0^+)\!+\!\frac1\gamma_p\!\int_0^t \!\!dt'\,
e^{-k_p(t-t')/\gamma_p}f_p(t')
\label{e13}
\end{eqnarray} 
for $\vec Y_p=[\vec Y^{(l)}_p,\vec Y^{(r)}_p]$, yielding
\begin{eqnarray}
\frac{\vec\phi(t)}k\!=\!\underbrace{-\frac8N\delta\vec
r_0\!\sum_p\!e^{-k_pt/\gamma_p}}_{\vec
q(t)/k}+\!\underbrace{4\sum_p\!\frac{\pi(2p+1)}N\vec g_p(t)}_{\vec
g(t)/k},\, \mbox{with}\nonumber\\ &&\hspace{-8.5cm}\vec
g_p(t)=\!e^{-k_pt/\gamma_p}[\vec Y^{(r)}_p(0^-)+\vec
Y^{(l)}_p(0^-)]\nonumber\\ &&\hspace{-7.5cm}+\frac1\gamma_p\int_0^t
\!\!dt'\, e^{-k_p(t-t')/\gamma_p}[f^{(r)}_p(t')+f^{(l)}_p(t')].
\label{e14}
\end{eqnarray} 
It is now seen, by converting the sum to an integral, that
\begin{eqnarray} 
\vec q(t)=-\frac{8k}N\vec{\delta r}_0\sum_p
e^{-k_pt/\gamma_p}=-2\vec{\delta r}_0\, \sqrt{\frac{\pi\gamma k}
{t}}\, e^{-t/\tau}.
\label{e15}
\end{eqnarray} 
It is also seen, using Eq. (\ref{e8}), that $\langle\vec
g(t)\rangle_0=0$, with the FDT
\begin{eqnarray} 
\langle\vec g(t)\cdot\vec
g(t')\rangle_0=\frac{24kk_BT}N\sum_p\!e^{-k_pt/\gamma_p}=6k_BT\sqrt{\frac{\pi\gamma
k}{(t-t')}}\, e^{-(t-t')/\tau}.
\label{e16}
\end{eqnarray} 
Here $\langle\ldots\rangle_0$ denotes the average over an equilibrated
ensemble of polymers at $t=0^-$; for which the middle monomer of each
polymer moves by $\vec{\delta r}_0$ at $t=0$.

The above procedure is trivially generalized to obtain the GLE
(\ref{egleb}): one needs to consider an ensemble of polymers that
moved by a distance $\vec{\delta r}_0$ at $t=0$, $\vec{\delta r}_1$ at
$t=t_1$, and so on. For this ensemble, having recognized that $\vec
v(t)=\sum_i\vec{\delta r}_i\,\delta(t-t_i)$ [with $t_0=0$], where the
angular brackets include an average over an equilibrated ensemble of
polymers at $t=0^-$, one arrives at Eq. (\ref{egleb}). Further, the
GLE (\ref{eglea}) is obtained by first Laplace transforming
Eq. (\ref{egleb}), then expressing the velocity of the middle monomer
in terms of the chain tension it experiences in the Laplace space, and
finally inverting the Laplace transform to return to real time,
resulting in the FDT $\langle\vec v(t)\cdot\vec
v(t')\rangle_{\vec\phi=0}=3k_BT\,a(t-t')\sim
(t-t')^{-3/2}e^{-(t-t')/\tau}$ \cite{laplace}. Subsequently, the
result that the MSD of the middle monomer increases $\sim t^{1/2}$
till time $\tau$ and $\sim t$ thereafter is obtained by integrating of
$\langle\vec v(t)\cdot\vec v(t')\rangle_{\vec\phi=0}$ twice in time.

At this point I make the important observation for the GLE formulation
of Eqs. (\ref{egleb}-\ref{eglea}) that if $\mu(t)\sim t^{-\alpha}$,
then the MSD of the middle monomer has to increase $\sim t^\alpha$.

{\it Polymeric response to local strains, and the GLE.\/} From
Eq. (\ref{egleb}) it is clear that $\mu(t)$ is the mean response of
the polymer to a local strain --- i.e., altered chain tension at the
middle monomer --- created by moving the middle monomer by a distance
$\vec{\delta r}$ at $t=0$ and fixing it at its new position $\forall
t$. The mean local strain then relaxes in time $\sim t^{-1/2}$, i.e.,
$\langle\vec\phi(t)\cdot\vec\phi(0)\rangle\sim t^{-1/2}$. While the
response of a polymer to a local strain depends on how the strain is
created, {\it the identification of $\mu(t)$ as the polymer's mean
local strain relaxation response alone allows one to write down the
GLE\/}, as I show below. First, given the identification of $\mu(t)$
as the polymer's mean local strain relaxation response one can always
write the stochastic Eq. (\ref{egleb}) with $\langle\vec
g(t)\rangle=0$, which holds by definition. Next, to obtain the FDT,
consider Eq. (\ref{egleb}) for an ensemble of polymers with $\vec
v(t)=0\,\,\forall t$ and $\vec\phi(t_0)=\vec g_0$, a specific
value. For such an ensemble $\vec g(t)\equiv\vec\phi(t)$, and since
$\mu(t)$ is the polymer's mean local strain relaxation response,
$\langle\vec\phi(t)\cdot\vec\phi(t_0)\rangle=g^2_0\,\mu(t-t_0)$ for
$t>t_0$. Extending this to the dynamics of a polymer in an equilibrium
ensemble (where $\vec g_0$ is also chosen from the equilibrium
ensemble), one has $\langle\vec g(t)\cdot\vec
g(t_0)\rangle\equiv\langle\vec\phi(t)\cdot\vec\phi(t')\rangle_{\vec
v=0}=\langle \phi^2(t)\rangle_{\vec v=0}\,\mu(t-t')$ [for a phantom
polymer $\langle \phi^2(t)\rangle_{\vec v=0}=3k_BT$].

Using the identification of $\mu(t)$ in Eq. (\ref{egleb}) as the mean
polymeric response to local strain leading to the GLE, I now argue the
existence of the GLE for self-avoiding polymers and for polymer
translocation.

{\it The GLE for self-avoiding polymers.}  The monomers of a
self-avoiding polymer interact over a long-range, which prohibits one
from writing down an exact equation for the velocities of the monomers
in terms of the forces they experience. However, quite a few
properties of self-avoiding polymers are well-known: two of them we
need here for a polymer of length $N$ are: (i) the terminal time
$\tau$ scales $\sim N^{1+2\nu}$ for a Rouse polymer, and as $\sim
N^{3\nu}$ for a Zimm polymer \cite{de}; and (ii) the entropic spring
constant of a polymer scales as $N^{-2\nu}$ \cite{degennes}. Here
$\nu$ is the Flory exponent, in three dimensions $\nu\approx0.588$,
and in two dimensions $\nu=0.75$. Imagine that one moves the middle
monomer of a self-avoiding polymer by a small distance $\vec{\delta
r}$ at $t=0$ and holds it at its new position, corresponding to
$\vec{v}(t)=\vec{\delta r}\,\delta(t)$. Following (i), at time $t$,
counting away from the middle monomer, all the monomers within a
backbone distance $n_t\sim t^{1/(1+2\nu)}$ for a Rouse, and $\sim
t^{1/(3\nu)}$ for a Zimm polymer equilibrate to the new position of
the middle monomer. However, since the rest $(N-n_t)$ monomers are not
equilibrated to the new position of the middle monomer at time $t$,
these $n_t$ monomers are stretched by a distance $\vec{\delta
r}$. With the entropic spring constant of these $n_t$ equilibrated
monomers scaling $\sim n_t^{-2\nu}$ [following (ii)], the mean force
the middle monomer will experience at its new position is given by
$\vec\phi(t)\sim n_t^{-2\nu}(-\vec{\delta r})\sim
t^{-2\nu/(1+2\nu)}(-\vec{\delta r})$ for a Rouse, and $\vec\phi(t)\sim
n_t^{-2\nu}(-\vec{\delta r})\sim t^{-2/3}(-\vec{\delta r})$ for a Zimm
polymer [force $=$ (spring constant) $\times$ (stretching
distance)]. This power-law behavior lasts only till the terminal time
$\tau$. [The time behavior of Eq. (\ref{e15}) is recovered from this
line of argument upon simply replacing $\nu$ by $1/2$ corresponding to
a phantom Rouse polymer.]  Such behavior of $\mu(t)$, in light of the
above paragraphs, implies that the motion of the middle monomer of the
Rouse and the Zimm polymers is indeed described by the GLE,
reproducing the well-known results that the MSD of the middle monomer
increases respectively $\sim t^{2\nu/(1+2\nu)}$ and $\sim t^{2/3}$,
and $\sim t$ thereafter. The GLE for a self-avoiding Rouse polymer has
recently been confirmed numerically \cite{rousepaper}.

{\it The GLE for polymer translocation.}  Polymer translocation is the
process where a polymer passes through a narrow pore in a membrane. Of
interest here is the so-called unbiased (i.e., in the absence of any
force or field) translocation: the polymer passes through the pore
purely due to thermal fluctuations, and the dynamics is anomalous
\cite{dprl}. A translocating polymer essentially consists of two
sub-polymers --- one on each side of the membrane --- exchanging
monomers through the pore. When a monomer translocates, the polymer's
chain tension at the pore changes: it increases on the side of the
membrane which the monomer comes from, and decreases on the other. The
relevant polymeric response therefore, is to a (local) strain due to
extra monomers injection into a tethered polymer at the tether
point. Consider the case where $n$ extra monomers are injected into a
tethered polymer at the tether point at $t=0$. For phantom Rouse
polymers the mean response to such a strain is given by $\mu(t)\sim
t^{-1}e^{-t/\tau}$, with $\tau\sim N^2$ \cite{dpre}. For self-avoiding
polymers $\mu(t)$ is obtained as follows. Following (i), at time $t$,
counting away from the pore, all the monomers within a backbone
distance $n_t\sim t^{1/(1+2\nu)}$ for a Rouse, and $\sim t^{1/(3\nu)}$
for a Zimm polymer, equilibrate to the injected monomers. The real
space extent of $n_t$ monomers is $r(n_{t})\sim n_t^\nu$, but since
the rest $(N-n_t)$ monomers are not equilibrated to the injected
monomers at time $t$, there are $(n_t+n)$ monomers squeezed in a space
that extends only to $r(n_{t})$. The corresponding compressive force
[force $=$ (spring constant) $\times$ (stretching distance)] from
these $(n_{t}+n)$ monomers, felt at the pore, and hence $\mu(t)$, is
the given by $\sim n_t^{-2\nu}[\delta r(n_{t})]\sim
n_t^{-2\nu}\,n[\partial r(n_{t})/\partial n_t]=\nu nn_{t}^{-(1+\nu)}$,
which scales $\sim t^{-(1+\nu)/(1+2\nu)}$ for a Rouse and $\sim
t^{-(1+\nu)/(3\nu)}$ for a Zimm polymer. (Once again, this behavior
lasts only till the terminal time $\tau$.) This implies that polymer
translocation is described by the GLE as well, resulting in the
scaling of the MSD $\sim t^{(1+\nu)/(1+2\nu)}$ for self-avoiding Rouse
and $\sim t^{(1+\nu)/(3\nu)}$ for self-avoiding Zimm polymers up to
time $\tau$ and $\sim t$ thereafter. Consequently, the pore-blockade
time scales $\sim N^2$ for a phantom Rouse \cite{muthu}, $\sim
N^{2+\nu}$ for self-avoiding Rouse \cite{anom1,anom2}, and $\sim
N^{1+2\nu}$ for a self-avoiding Zimm polymer \cite{anom1}.

In summary, in view of the recent popularity of the GLE to reproduce
the anomalous dynamics in some stochastic dynamical systems, in this
Letter I have concerned myself with the question whether a system with
anomalous dynamics can lead to the GLE formulation, and have
considered polymer physics, where sub-diffusive behavior is
commonplace. I have provided an exact derivation of the GLE for
phantom Rouse polymers, and have argued the case for the (necessary)
existence of the GLE for self-avoiding polymers and for polymer
translocation. The key to show the existence of the GLE for these
cases is to relate the (monomeric) velocity to the power-law mean
(polymeric) response behavior of to local strains. This implies that
there exists the GLE formulation for any system (that exhibits
anomalous dynamics) for which the velocity gives rise to local
strains, and the local strain relaxes in a power-law fashion.
Moreover, all the cases considered here concern single polymer
dynamics; whether the principle holds for many-polymer systems, e.g., 
for subdiffusive behavior in polymer melts, or for superdiffusive 
behavior in ``living polymers'' \cite{langb} remains to be 
investigated.


\begin{thebibliography}{09}
\bibitem{laplace} Note that $\tilde\mu(s)\tilde a(s)=1$ yields $\tilde
a(s)\sim(s+\tau^{-1})^{1/2}$, the inverse Laplace transform of which,
strictly speaking, does not exist. This can be traced back to
Eq. (\ref{e15}), where I have ignored the minimum distance between the
consecutive monomers providing the ultraviolet cutoff for $p$ for the
summation. In reality, the $t^{-1/2}$ behavior does not extend all the
way to $t=0$, implying that the inverse Laplace transform of $\tilde
a(s)$ is indeed given by $t^{-3/2}$.
\end{thebibliography}
\end{document}